\documentclass[proceedings]{JHEP37}
\usepackage{eurosym}
\usepackage{amssymb}
\usepackage{amsfonts,cite}
\usepackage{amsmath}
\usepackage{graphicx}
\usepackage{multirow}
\usepackage{epsfig}

\DeclareMathOperator\sech{sech}

\newbox\mybox

\newcommand\fverb{\setbox\mybox=\hbox\bgroup\verb}
\newcommand\fverbdo{\egroup\medskip\noindent\fbox{\unhbox\mybox}\ }
\newcommand\fverbit{\egroup\item[\fbox{\unhbox\mybox}]}
\conference{Stability in integrable nonlocal nonlinear equations}
\abstract{Recently a variety of nonlocal integrable systems has been introduced that besides fields located at particlar space-time points simultaneously also contain fields that are located at different, but symmetrically related, points. Here we investigate different types of soliton solutions with regard to their stability against linear pertubations obtained for the nonlocal version of the Hirota/nonlinear Schr\"odinger equation and the so-called Alice and Bob versions of the Korteweg-de Vries and Bousinesq equations. We encounter different types of scenarios: Solition solutions that are linearly stable or unstable and also solutions that change their stability properties depending on the parameter regime they are in.}  

\title{Stability in integrable nonlocal nonlinear equations}
\author{Julia Cen$^\ast$, Francisco Correa$^\circ$, Andreas Fring$^\bullet$ and Takanobu Taira$%
^\bullet$ \\
$\ast$ Theoretical Division and Center for Nonlinear Studies, Los Alamos National\\$\,\,$  Laboratory, Los Alamos, New Mexico 87545, USA\\
$\circ$ Instituto de Ciencias F{\'{\i}}sicas y Matem{\'{a}}ticas,
Universidad Austral de Chile, \\
$\,\,$ Casilla 567, Valdivia, Chile\\
$\bullet$ Department of Mathematics, City, University of London,\\
$\,\,$ Northampton Square, London EC1V 0HB, UK \\

E-mail: julia.cen@outlook.com, francisco.correa@uach.cl, a.fring@city.ac.uk,
takanobu.taira@city.ac.uk}
\input{tcilatex}
\begin{document}

\section{Introduction}
Quantum nonlocality is a well-established phenomenon \cite{aspect1982} that does not allow for an entirely local description, but must in some form take correlations between events at different space-time points into account. In the context of classical integrable field theories it was recently observed  \cite{ablowitz2013} that when exploiting different versions of ${\cal CPT}$-symmetries, theories containing fields at non-identical space-time points naturally arise. Consistent multi-soliton solutions to these type of models have been constructed for nonlinear Sch\"odinger equations \cite{ablowitz2013}, Hirota systems \cite{CenFringHir,li2021inverse} and so-called Alice and Bob versions  of the Korteweg-de Vries equation (AB-KdV) \cite{AliceB1,AliceB2,AliceB3}, Boussinesq equation (AB-Boussinesq) \cite{li2019multiple} and other systems. As a concrete potential physical application it was shown for instance in \cite{AliceB2} that a particular version of AB-KdV systems can be derived as a multiple vorticity interaction model related to a standard atmospheric and oceanic dynamical system.     

As these novel systems allow for different types of solutions one needs to specify under which conditions they are considered to be meaningful. For solutions to be physically relevant one naturally requires their energies, possibly also other conserved quantities that have physical interpretations, to be real and in addition demands them to be stable. Despite some of the solutions to be complex, the reality of their energies is well investigated and can be established by making use of general arguments based on the underlying ${\cal{CPT}}$-symmetry of the models \cite{AFKdV,CenFring,fring2020BPS,CFTSkyrmions}. However, the crucial stability property has been largely ignored so far and the main purpose of this manuscript is to start filling this gap.

Our manuscript is organised as follows: In section 2 we discuss the stability properties of two types of previously constructed one-soliton solutions for the Hirota equation. In section 3 and 4 we discuss the stability of Alice and Bob KdV and Boussinesq systems, respectively. Our conclusions are drawn in section 5.

\section{Nonlocal Hirota equation with parity conjugated fields}

Let us start by recalling from \cite{CenFringHir} the nonlocal Hirota equation that besides fields $q(x,t)$ also contains their parity conjugate fields $q^\ast(-x,t)$ 
\begin{equation}
	\!iq_{t} =-\alpha \left( q_{xx}-2\kappa \tilde{q}^{\ast }q^{2}\right)
	\!+\delta \left( q_{xxx}-6\kappa q\tilde{q}^{\ast }q_{x}\right) , \qquad \alpha, \delta \in \mathbb{R}, \kappa = \pm 1
	\label{new1}
\end{equation}
where we use the abbreviations $q:=q(x,t)$, $\tilde{q}:=q(-x,t)$ with $q^\ast, \tilde{q}^\ast$ denoting their respective complex conjugates. Since this equation contains fields located at $x$ as well as fields situated at $-x$, these type of equations are referred to as {\em nonlocal}, as opposed to {\em local} when all fields are situated at the same particular point $(x,t)$. In the limit $\delta \rightarrow 0$ we recover the nonlocal version of the nonlinear Schr\"odinger equation, first proposed in \cite{ablowitz2013}. The nonlocal versions of well-known integrable systems were derived by introducing a different type compatibility procedure for the two equations that emerge in a Lax pair or AKNS construction. Instead of making the standard choices for certain quantities so that the two equations are simply related by complex conjugation, one may also consistently relate them by a parity, time, parity/time conjugation or simply by parity, time, parity/time transformation. In this section we exclusively focus on the version that relates the equations by parity conjugation.

Alternatively, we demonstrate here that the nonlocal Hirota equation (\ref{new1}) can also be derived directly in a conventional fashion from a Hamiltonian $H =\int {\cal{H}} dx$ with complex parity invariant Hamiltonian density ${\cal{H}} $. We recall for this purpose the lowest nonlocal charges ${\mathcal I}_n = \int w_n  dx$ with charge densities $w_n$ for the nonlocal Hirota system that were systematically derived in \cite{CenFringHir}. The lowest charges read 
\begin{equation}
	w_0= q \tilde{q}^\ast, \quad 
	w_1=- q \tilde{q}^\ast_x, \quad
	w_2= q \tilde{q}^\ast_{xx}-q^2 \left(\tilde{q}^\ast \right)^2, \quad
	w_3= q q_x (\tilde{q}^\ast)^2 + 4 q^2 \tilde{q}^\ast \tilde{q}^\ast_x- q \tilde{q}^\ast_{xxx} .
 \label{charges}
\end{equation}
The complex parity transformed quantities $  \tilde{{\cal I}}^\ast_n =\int  \tilde{w}^\ast_n  dx$ are also conserved. Using these charge densities we define now the complex parity invariant Hamiltonian density
\begin{eqnarray}
 {\cal{H}} &=& -\frac{\alpha}{2} (w_2 + \tilde{w}^\ast_2) + \frac{\delta}{2} (w_3 + \tilde{w}^\ast_3) + \frac{\kappa-1}{2}\left[  2 \alpha w_0^2 - 3 \delta w_0 (w_1 + \tilde{w}^\ast_1)    \right]    \\
 & =&\frac{\alpha}{2} \left[ 2 \kappa q^2 (\tilde{q}^\ast)^2 -(q \tilde{q}^\ast_{xx} + \tilde{q}^\ast q_{xx}   )        \right]+ \frac{\delta}{2} \left[  3 \kappa \left( q^2 \tilde{q}^\ast \tilde{q}^\ast_x- (\tilde{q}^\ast)^2 q q_x  \right)  -(q \tilde{q}^\ast_{xxx} - \tilde{q}^\ast q_{xxx}   )  \right], \qquad
	\label{Ham1}
\end{eqnarray}
which by using functional variations for complex Hamiltonians in the form
\begin{equation}
	i q_t = \frac{\delta H }{\delta \tilde{q}^\ast} = \frac{\partial {\cal{H}}}{ \partial \tilde{q}^\ast} - \frac{d}{dx} \frac{\partial {\cal{H}}}{ \partial \tilde{q}^\ast_x}
	+ \frac{d^2}{dx^2} \frac{\partial {\cal{H}}}{ \partial \tilde{q}^\ast_{xx}}+ \ldots
\end{equation}
leads to the nonlocal Hirota equation (\ref{new1}). The complex parity conjugate equation of (\ref{new1}) is then simply obtained from 
\begin{equation}
	-i \tilde{q}^\ast_t = \frac{\delta H }{\delta q} = \frac{\partial {\cal{H}}}{ \partial q} - \frac{d}{dx} \frac{\partial {\cal{H}}}{ \partial q_x}
	+ \frac{d^2}{dx^2} \frac{\partial {\cal{H}}}{ \partial q_{xx}}+ \ldots
\end{equation} 
The Hamiltonian allows for a conventional interpretation and in particular enables us to compute the energy of particular solutions. We note that all terms in the Hamiltonian are nonlocal involving interactions between fields and their derivatives at $x$ and $-x$.

For physically meaningful solutions we demand the charges, and in particular the energy measured by the Hamiltonian, to be real.

\subsection{Linear stability of nonlocal solutions}
We now address the central question of the manuscript and explore whether a particular solution $q_0$ of (\ref{new1}) is stable or unstable. There exist a large variety of notions of stability properties and techniques to study them. Here we investigate their linear stability by carrying out an analysis that was previously employed to establish the stability of solutions to the nonlinear Schr\"odinger equation \cite{herbst1985stability}. Surprisingly it will turn out that the nonlocality of the problem simplifies the techniques involved. 

To start we perturb a solution $q_0$ in the usual fashion by replacing
\begin{equation}
	q(x,t) \rightarrow  	q_0(x,t) + \epsilon \sigma(x,t), \qquad \epsilon \ll 1, \label{zeropert}
\end{equation}
and seek the properties of the perturbing function $\sigma(x,t)$, in particular its behaviour in time. We recover (\ref{new1}) for $q_0$ in zeroth order and to establish the stability of $q_0$ we demand in addition that the first order terms in $\epsilon$ vanish
\begin{equation}
    i \sigma _t+\alpha  \left(\sigma_{xx}-2 \kappa  q_0^2 \tilde{\sigma }^*-4 \kappa  q_0 \sigma  \tilde{q}_0^*\right) + \delta  \left\{ 6 \kappa \left[ q_0 \tilde{\sigma }^* (q_0)_x+  q_0  \tilde{q}_0^* \sigma _x+  \sigma 
    \tilde{q}_0^* (q_0)_x\right] -\sigma _{xxx}\right\}=0. \label{pert1}
\end{equation}
Following the standard interpretation, we infer that in the scenario when solutions $\sigma(x,t)$ to (\ref{pert1}) can be found that only introduce oscillations or deformations onto the original solution, $q_0$ is regarded as stable, whereas when $\sigma(x,t)$  diverges as a function of time then the original solution is unstable.  

  To solve the first order auxiliary equation (\ref{pert1}) we are now making a few modifications and assumptions. First of all we replace $x,t$ by a new set of variables $z,\bar{z}$ defined as
\begin{equation}
z: =x- i \delta \mu^2  t, \qquad \text{and} \qquad   \bar{z} :=x+ i \delta \mu^2 t,
\end{equation}
so that
\begin{equation}
  x= \frac{z+\bar{z}}{2}, \qquad	t=\frac{i}{2 \delta \mu^2 } (z-\bar{z}), \qquad \partial_x= \partial_z + \partial_{\bar{z}}, \qquad \partial_t = i \delta \mu^2 (\partial_{\bar{z}} - \partial_z ).
\end{equation}
Next we assume the original solution and the perturbing function to factorize in the form 
\begin{equation}
q_0(x,t)=\mu e^{i  \alpha \mu^2 t} u[\mu z(x,t)], \qquad \sigma(x,t)=i a e^{i \alpha \mu^2 t} \psi[\mu z(x,t)], \label{q0}
\end{equation}
where $a \in \mathbb{R}$ and $u$, $\psi$ are functions that only depend on $z$ and not $\bar{z}$. The parity conjugates of these functions are assumed to be of the form
\begin{equation}
	q_0^\ast(-x,t)=\mu e^{-i t \alpha \mu^2 t} u[\mu z(x,t)], \qquad \sigma^\ast(-x,t)=-i a e^{-i t \alpha \mu^2 t} \omega \psi[\mu z(x,t)],
\end{equation}
where $\omega=\pm 1$ allows for $\psi$ to be complex parity even or odd. When substituting these functions into the first order auxiliary equation (\ref{pert1}) we obtain
\begin{equation}
	i a \mu ^2   e^{\frac{\alpha  (\bar{z}-z)}{2 \delta }} \left(\alpha \, \text{Aux}_1+  \delta \,\text{Aux}_2  \right) =0,
\end{equation}
where
\begin{eqnarray}
	\text{Aux}_1 &=&  \left[2 \kappa  (\omega -2) u^2(\mu  z)-1\right] \psi (\mu  z) +\psi ''(\mu  z),   \\
	\text{Aux}_2 &=& \mu  \left\{6 \kappa  (1-\omega ) u(\mu  z)  u'(\mu  z)\psi (\mu  z)  +\left[ 1+6 \kappa  u(\mu  z)^2 \right] \psi '(\mu  z)-\psi ^{(3)}(\mu  z) \right\}. \qquad
\end{eqnarray}
We need to distinguish now between complex parity even and odd functions $\psi$. For $\omega=-1$ we find that
\begin{equation}
	\partial_z \text{Aux}_1 = - \text{Aux}_2 ,
\end{equation}
so that in this case the problem of solving  (\ref{pert1}) is reduced to the Sturm-Liouville problem
\begin{equation}
   \psi ''(\mu  z)- 6 \kappa   u^2(\mu  z) \psi (\mu  z)  =\psi (\mu  z), \label{SL}
\end{equation}
for the potential $V=- 6 \kappa   u^2$. In contrast, when $\omega=1$ we obtain that
\begin{equation}
	\partial_z \text{Aux}_1 = - \text{Aux}_2 + 4 \kappa \mu u(\mu z) W(u,\psi) ,
\end{equation}
where $W(u,\psi)=u \psi'-u' \psi $ is the Wronskian for the functions $u$ and $\psi$. This equation may be solved by taking $\psi (\mu  z) = u (\mu  z)$, so that the Wronskian vanishes, and when in addition $u (\mu  z)$ solves the equation  
\begin{equation}
	u ''(\mu  z)- 2 \kappa   u^3(\mu  z)  =u (\mu  z) . \label{uconst}
\end{equation}
Let us now see whether we can find solutions $q_0$ that are of the specified form and subsequently use the above reasoning to solve the first order equation (\ref{pert1}).

\subsection{One-soliton solutions}

In \cite{CenFringHir} two different types of N-soliton solutions to the nonlocal Hirota equation (\ref{new1}) with qualitatively different behaviour were derived using Hirota's direct method and also Darboux-Crum transformation. The one-soliton solutions read
\begin{equation}
	q_{\text{st}}^{(1)}=\frac{\lambda (\mu -\mu ^{\ast })^{2}\tau _{\mu ,\gamma }%
	}{(\mu -\mu ^{\ast })^{2}+\left\vert \lambda \right\vert ^{2}\tau _{\mu
			,\gamma }\tilde{\tau}_{\mu ,\gamma }^{\ast }}, \qquad \text{and} \qquad
		q_{\text{nonst}}^{(1)}=\frac{(\mu +\nu )\tau _{\mu ,i\gamma }}{1+\tau _{\mu
				,i\gamma }\tilde{\tau}_{-\nu ,-i\theta }^{\ast }},
		 \label{sol2}
\end{equation}  
with $\lambda, \mu, \gamma \in \mathbb{C}$ and $\nu, \mu, \gamma, \theta \in \mathbb{R}$, respectively. The so-called tau-functions are defined as
\begin{equation}
	\tau _{\mu ,\gamma }(x,t):=e^{\mu x+\mu ^{2}(i\alpha -i \delta \mu )t+\gamma },
\end{equation}
and $\kappa$ is set to -1. More recently Li and Tian \cite{li2021inverse} recovered the nonstandard solutions $q_{\text{nonst}}^{(N)}$ using the inverse scattering method. This is reassuring, but contrary to the claim made in \cite{li2021inverse}, where the authors assert that their solutions were in fact new and different from those found in \cite{CenFringHir}. It is easily verified that the one-soliton solution (7.15) therein converts precisely into $q_{\text{nonst}}^{(1)}$ when identifying the parameters used in \cite{li2021inverse} simply as $\beta \rightarrow \delta$, $\eta_1 \rightarrow  \mu/2$,  $\bar{\eta}_1 \rightarrow  -\nu/2$, $\theta_1 \rightarrow  -\gamma + \pi$ and $\bar{\theta}_1 \rightarrow  -\theta$.      

As discussed in \cite{CenFringHir} the behaviour of the two types of solutions in (\ref{sol2}) is quite different. While $q_{\text{st}}^{(1)}$ exhibits in general an oscillatory scattering type of behaviour, the solution $q_{\text{nonst}}^{(1)}$ can be of rogue wave type by tending to infinity at certain times $t_s$. Simply from the singularity structure of the solutions it is obvious that in those cases any kind of initial condition will blow up when evolved according to $q_{\text{nonst}}^{(1)}$. This type of behaviour was also observed for the nonlinear Schr\"odinger equation, i.e., $\delta=0$,  in \cite{ablowitz2013} and can be verified more formally using functional analytical methods \cite{genoud2017}. 

Here we are interested in the stability of these solutions when perturbed by additional fields. For simplicity we first focus on two special solutions in which some parameter choices are made without loosing the ability to display the different types of behaviours using the remaining variables. By setting $\mu \rightarrow i \mu$, $\gamma \rightarrow 0$, $\lambda \rightarrow -2 i \mu$ in $	q_{\text{st}}^{(1)}$ and $\gamma=\theta \rightarrow 0$, $\nu \rightarrow \mu$ in $q_{\text{nonst}}^{(1)}$ we obtain the two simpler variants
\begin{equation}
 q_{\text{st,s}}^{(1)}= \mu e^{-i \alpha  \mu ^2 t} \csc \left[\mu  \left(x+i \delta  \mu ^2 t\right)\right], \quad	\text{and} \quad  q_{\text{nonst,s}}^{(1)}=\mu  e^{i \alpha  \mu ^2 t} \sech \left[\mu  \left(x-i \delta  \mu ^2 t\right)\right]. \label{simplesol}
\end{equation}
We notice that for $x=0$ the solution $q_{\text{st,s}}^{(1)}$ is singular at the origin $t_s=0$, whereas $q_{\text{nonst,s}}^{(1)}$ is singular at the times $t_s= (4 n+1)\pi/2\delta \mu^3 $, $n \in \mathbb{Z}$. Thus the latter solution exhibits a rogue wave behaviour for $\mu \in \mathbb{R}$, but is regular in its entire domain when reduced to a solution for the nonlocal nonlinear Schr\"odinger equation when $\delta=0$.

\subsubsection{The nonstandard rogue wave solution}
\begin{figure}[h]
	\centering         
	\begin{minipage}[b]{0.52\textwidth}           \includegraphics[width=\textwidth]{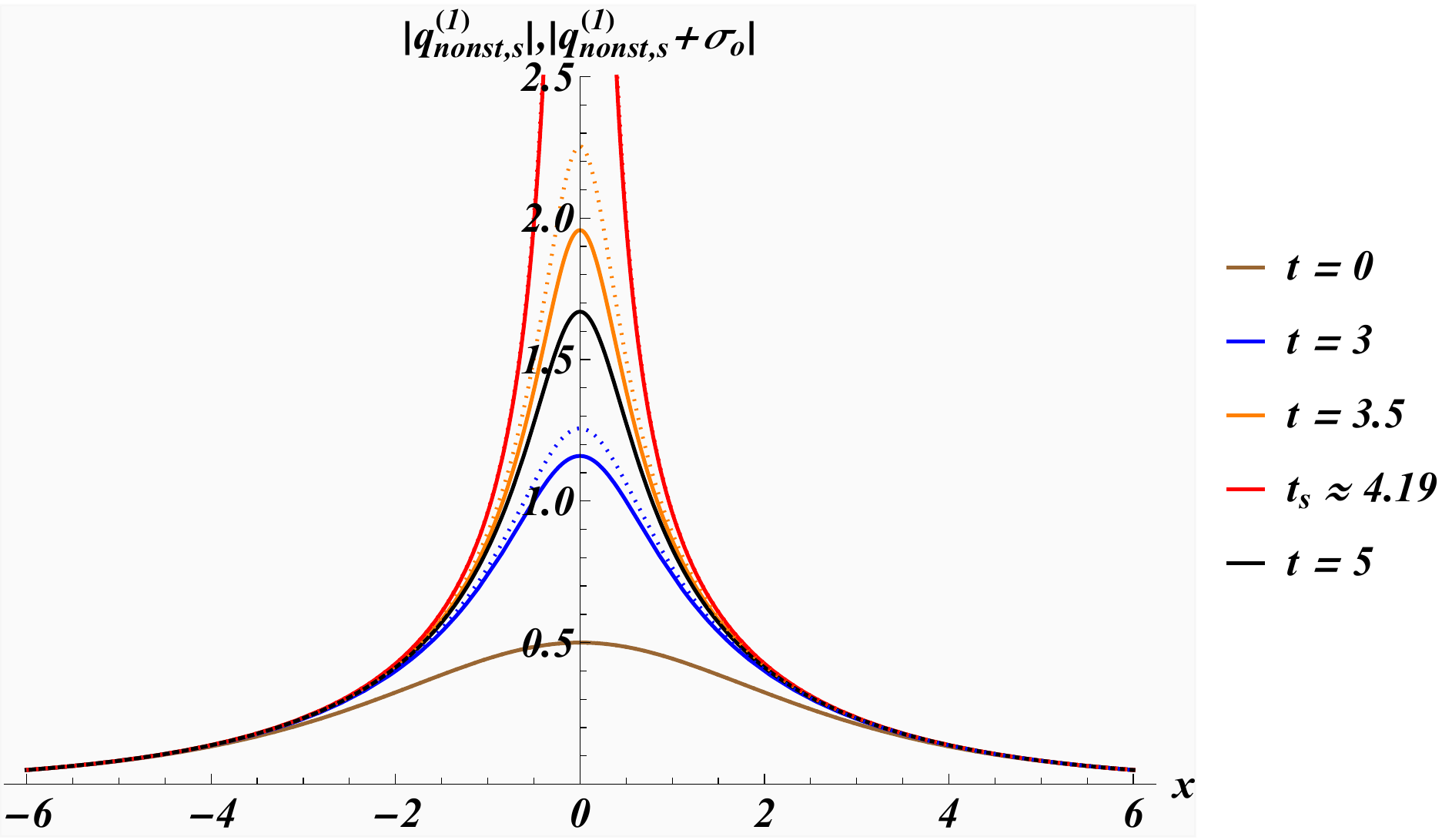}
	\end{minipage}   
	\begin{minipage}[b]{0.44\textwidth}           
		\includegraphics[width=\textwidth]{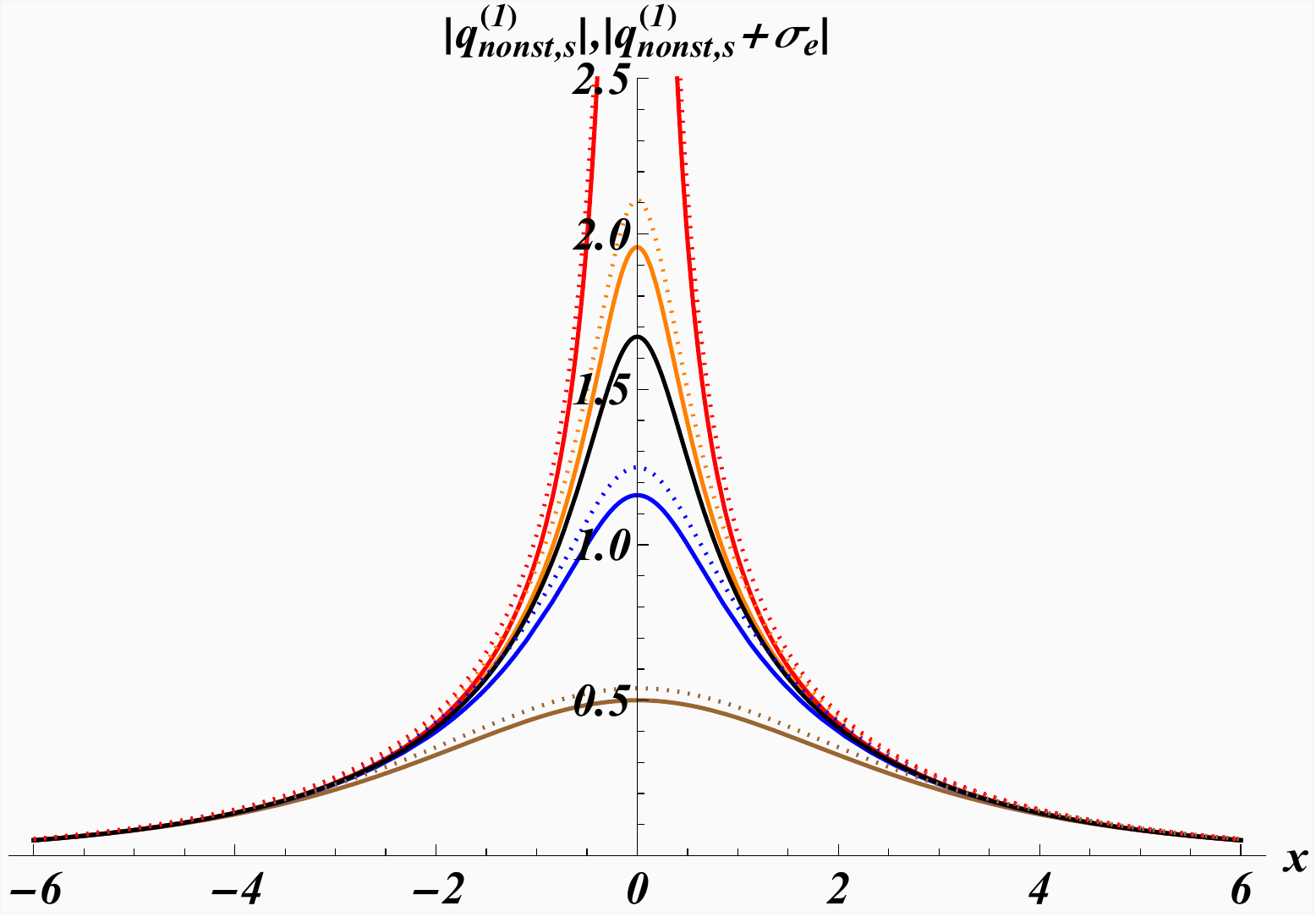}
	\end{minipage}    
	\caption{Rogue wave nonstandard one-soliton solution (\ref{sol2}) (solid) with its odd (\ref{perts1}) and even (\ref{perts2}) perturbed versions (dotted) at different times for $\epsilon = 0.02$ (odd), $\epsilon = 0.2$ (even), $\mu = 0.5 $, $a=1$, $\alpha=0.2$ and $\delta=3$.   }
	\label{nonspert}
\end{figure}

We observe now that the solution $q_{\text{nonst,s}}^{(1)}$ is a special case of the general form  $q_0$ as specified in equation (\ref{q0}) when identifying $u(\mu z) = \sech(\mu z) $. Solving for the odd case ($\omega=-1$) the Sturm-Liouville equation (\ref{SL}) with the corresponding ubiquitous\footnote{We recall that this potential with different types of scaling and overall shift also appears in the stability analysis of the $\phi^4$-theory, the sine-Gordon model \cite{jackiw1977q}, the Bullough-Dodd model \cite{BDstable} and the KdV equation as seen below. } $\sech^2$-potential, we obtain $\psi(\mu z)= \sech(\mu z) \tanh(\mu z)$, so that the first order equation (\ref{pert1}) is solved by 
\begin{equation}
\sigma_o(x,t) = i a e^{i  \alpha \mu^2 t} \sech\left[\mu (x- i \delta \mu^2  t) \right] \tanh\left[\mu (x- i \delta \mu^2  t) \right] .   \label{perts1}
\end{equation}
For the even case ($\omega=1$) we have to take $\psi(\mu z)=u(\mu z) = \sech(\mu z)$ and verify that $u(\mu z)$ indeed solves (\ref{uconst}). Thus we obtain the additional solution
\begin{equation}
	\sigma_e(x,t) = i a e^{i \alpha \mu^2 t} \sech\left[\mu (x- i \delta \mu^2  t) \right]  . \label{perts2}
\end{equation}
Both perturbations are asymptotically vanishing for $x \rightarrow \pm \infty $ and only introduce small variations of the solution as illustrated in figure \ref{nonspert}. We observe that the odd perturbations $\sigma_o(x,t)$ are more sensitive in distorting the solutions than the even perturbations $\sigma_e(x,t)$.  

One may easily verify that both solutions (\ref{perts1}) and (\ref{perts2}) are indeed genuine perturbations, and not symmetries of the original equation as the combination (\ref{zeropert}) is not a solution of the original equation.
We conclude that the nonlocal solution $q_{\text{nonst,s}}^{(1)}$ is stable with regard to the perturbations (\ref{perts1}) and (\ref{perts2}). 

It is worth noting that when comparing our stability analysis to the one carried out in the same spirit for the local nonlinear Schr\"odinger equation in \cite{herbst1985stability}, the Ansatz used here for the nonlocal case appears to be simpler. Whereas for the local case it was essential to take $\sigma(x,t)$ as a superposition of two separable functions coupled to each other in a nontrivial fashion, here when setting $\delta \rightarrow 0$ our Ansatz in (\ref{q0}) demonstrates that for the nonlocal case one may take the perturbing function as a simple factorization of one term only, as $\sigma(x,t)=f(x)g(t)$.

Besides being stable, we also find that the nonlocal charges (\ref{charges}) defined on $x \in (-\infty, \infty)$ obtained from these solutions are indeed real
\begin{equation}
	{\mathcal I}_0(q_{\text{nonst,s}}^{(1)}) =   2 |\mu|, \quad
	{\mathcal I}_1(q_{\text{nonst,s}}^{(1)}) = 0, \quad
	{\mathcal I}_2(q_{\text{nonst,s}}^{(1)}) = -  2|\mu| ^3, 
\end{equation}
including especially the energy $H=E$
\begin{equation}
E(q_{\text{nonst,s}}^{(1)}) = -\frac{2}{3}   \alpha  |\mu| ^3  .
\end{equation}
We notice that the energy of this solution is the same in the Hirota and the nonlinear Schr\"odinger equation, as there is no dependence on $\delta$.

The energies of the perturbed solutions are still constant in time, but acquire  correction terms
\begin{eqnarray}
	E(q_{\text{nonst,s}}^{(1)} + \sigma_o) &=& E(q_{\text{nonst,s}}^{(1)}) + \alpha  \left(\frac{2}{3} |\mu|  \epsilon ^2 -\frac{4 }{35 |\mu| } \epsilon ^4 \right), \qquad  \\
 E(q_{\text{nonst,s}}^{(1)} + \sigma_e) &=& E(q_{\text{nonst,s}}^{(1)})-  \alpha \left(2 |\mu|    \epsilon ^2+ \frac{4   }{3 |\mu| } \epsilon ^4 \right),
\end{eqnarray} 
where we identified the constant $a$ with $\epsilon$. Thus up to first order in $\epsilon$, the energies of the nonstandard solution and their perturbed version are identical. 

\subsubsection{The standard oscillatory solution} 

\begin{figure}[h]
	\centering         
	\begin{minipage}[b]{0.52\textwidth}           \includegraphics[width=\textwidth]{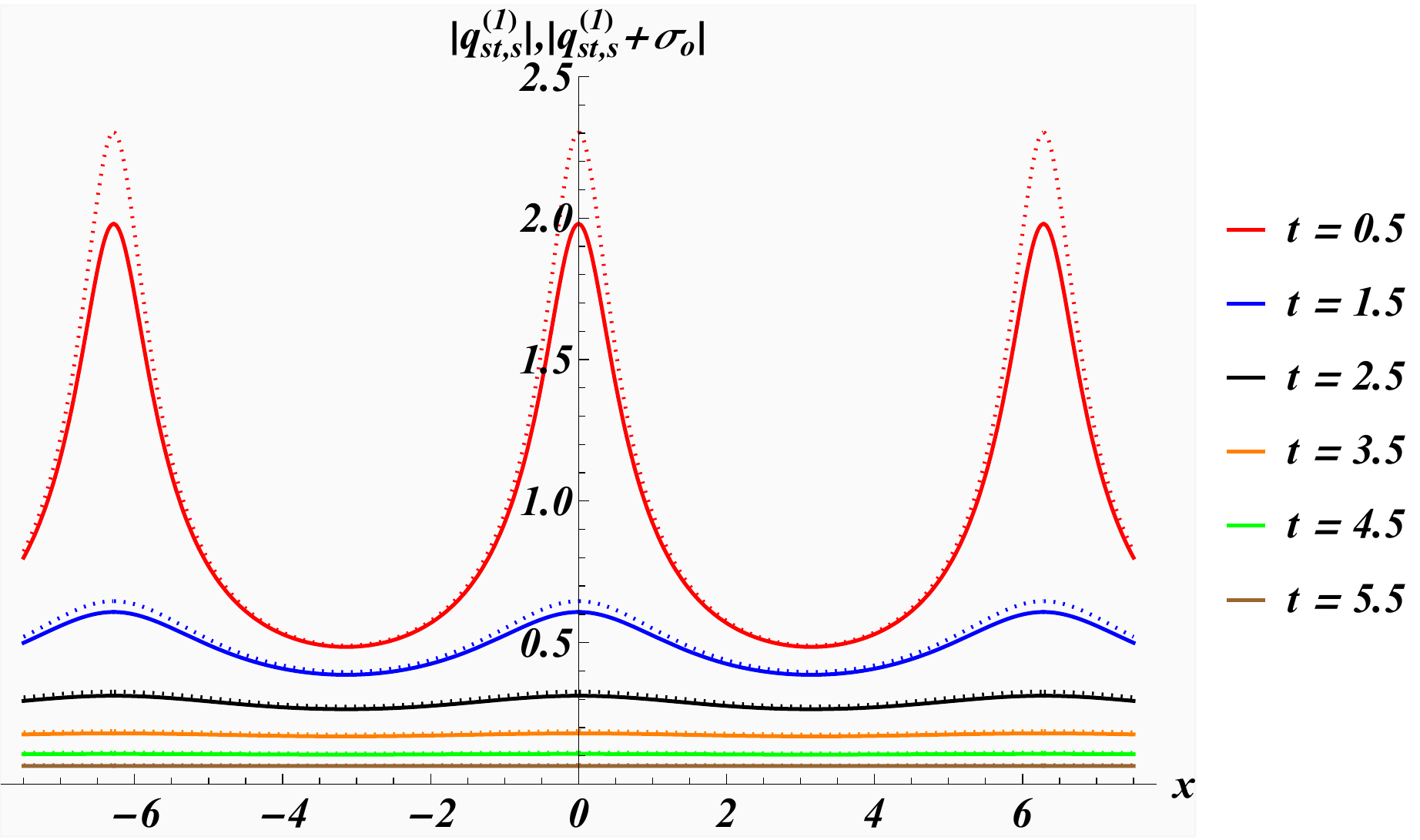}
	\end{minipage}   
	\begin{minipage}[b]{0.44\textwidth}           
		\includegraphics[width=\textwidth]{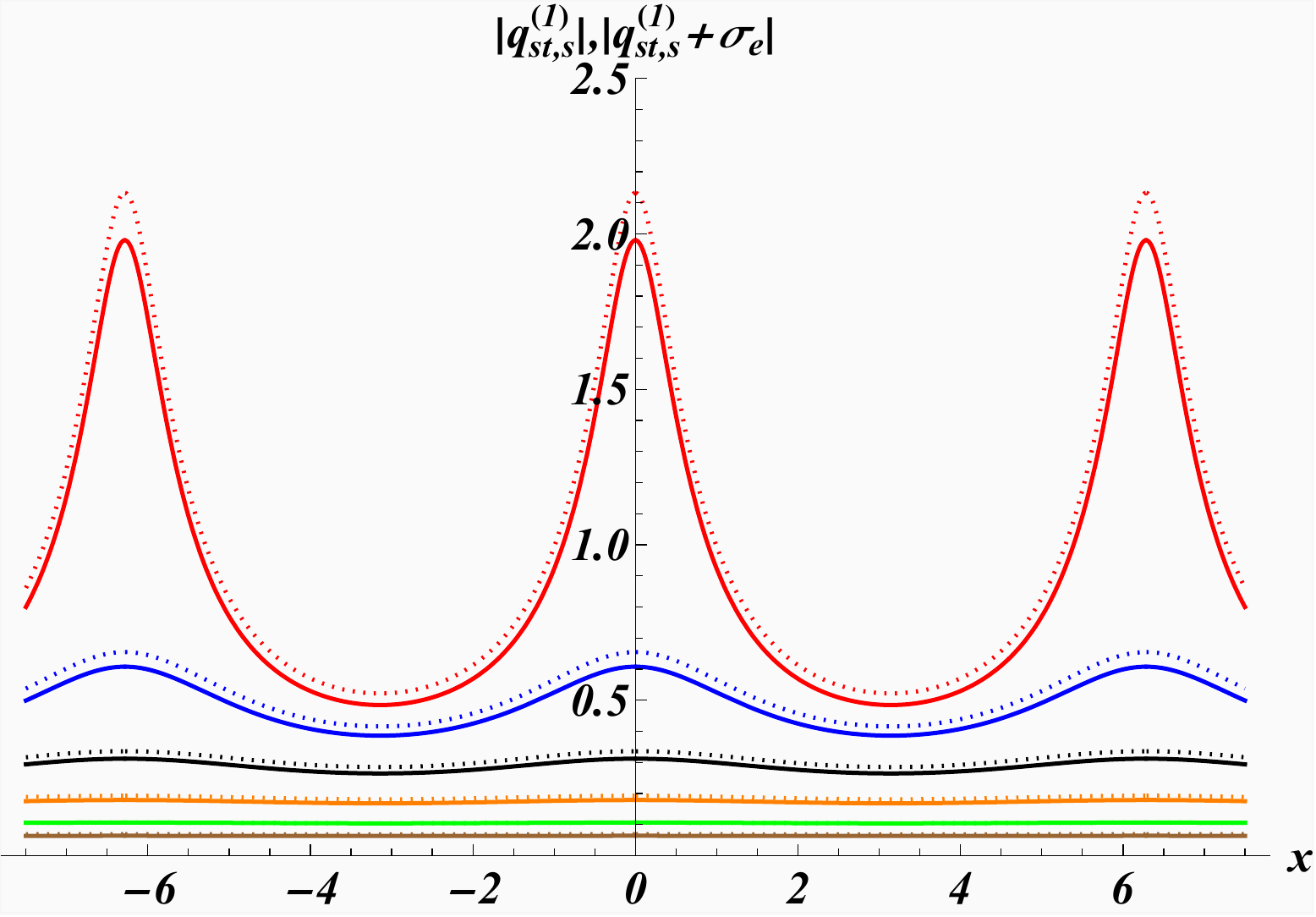}
	\end{minipage}    
	\caption{Decaying oscillatory standard one-soliton solution (\ref{sol2}) (solid) with its odd and even perturbed versions (dotted) at different times for $\epsilon = 0.02$ (odd), $\epsilon = 0.2$ (even),  $\mu = -0.5 $, $\alpha=0.3$ and $\delta=4$. }
	\label{Stsoloddeven}
\end{figure}

Noticing that we may convert the simplified versions of the solutions in (\ref{simplesol}) into each other as
\begin{equation}
	i q_{\text{nonst,s}}^{(1)} \left[ \mu \rightarrow i \mu , x \rightarrow x- i \pi /2 \mu   \right] =  q_{\text{st,s}}^{(1)} ,
\end{equation}
the computation for the standard solution goes along the same lines as for the nonstandard one. Thus it suffices to report the results. The auxiliary equation becomes a Sturm-Liouville equation for the solvable $csc^2$-potential from which we construct the perturbative terms as
\begin{eqnarray}
  \sigma_o(x,t) &=& i  e^{-i  \alpha \mu^2 t} \csc \left[\mu (x+ i \delta \mu^2  t) \right] \cot\left[\mu (x+ i \delta \mu^2  t) \right] ,\\
  \sigma_e(x,t) &=& i  e^{-i  \alpha \mu^2 t} \csc \left[\mu (x+ i \delta \mu^2  t) \right]  .
\end{eqnarray}
Thus as time evolves, the solutions for real values of $\mu$ as well as the perturbations tend to zero or infinity, but are both in sync as is seen in figure \ref{Stsoloddeven}. Similar to the nonstandard solution, also the standard solution is more sensitive to the even perturbations.  
For these standard solutions we need to compute the charges related to the densities (\ref{charges}) on a finite interval, e.g. from $-\pi/\mu$ to $\pi/\mu$. Proceeding in this manner we find that all charges, including their perturbations, are vanishing for $t \neq 0$ and $\delta \neq 0$.

\section{Alice and Bob KdV system}

Another closely related and well-studied class of nonlocal systems can be constructed from the prototype nonlinear wave equation, the Korteweg-de Vries (KdV) \cite{KdV} equation  
\begin{equation}
	u_{t}+6uu_{x}+u_{xxx}=0.  \label{KdV}
\end{equation}
The stability of the KdV one-soliton \cite{Ben} and multi-soliton solutions \cite{Sachs} has been established formally some time ago. Linear stability as previously discussed is easily seen to hold. For the investigation of some special type of solutions we may use the more convenient travelling wave coordinate $\zeta:= \left(\alpha  x-\alpha ^3 t\right)/2 $ and convert (\ref{KdV}) into the ordinary differential equation
\begin{equation}
	u_{\zeta \zeta \zeta}+\frac{24}{\alpha^2} uu_{\zeta}- 4 u_{\zeta}=0.  \label{KdVt}
\end{equation}
When replacing
\begin{equation}
	u(\zeta) \rightarrow  	u_0(\zeta) + \epsilon \sigma_u(\zeta), \qquad \epsilon \ll 1, 
\end{equation}
in (\ref{KdVt}) the equation of first order in $\epsilon$ becomes
\begin{equation}
   \partial_\zeta \left[	(\sigma_u)_{\zeta \zeta}+\frac{24   u_0}{\alpha ^2}\sigma_u -4 \sigma_u  \right] =0 .
\end{equation}
For the well-known KdV one-soliton solution 
\begin{equation}
	u_0(x,t) = \frac{\alpha ^2}{2}  \sech^2\left[\frac{1}{2} \left(\alpha  x-\alpha ^3 t\right)\right], \label{ones}
\end{equation}
the associated Sturm-Liouville auxiliary problem in the square bracket is easily solved to
\begin{equation}
	\sigma_u(x,t) = \sech^2\left[\frac{1}{2} \left(\alpha  x-\alpha ^3 t\right)\right] \tanh \left[\frac{1}{2} \left(\alpha  x-\alpha ^3 t\right)\right] .
\end{equation}

A nonlocal version of the KdV equation can be obtained from its Lax pair \cite{AliceB1,AliceB2,AliceB3}.
An alternative more direct way of construction consists of first decomposing the KdV field as $u(x,t)=1/2\left[ a(x,t)+b(x,t)\right] $, where
the two components are $\mathcal{PT}$-symmetrically  related as $\mathcal{PT}%
a(x,t)=a(-x,-t)=b(x,t)$ \cite{AliceB1,AliceB2,AliceB3}. In this way equation (\ref{KdV}) converts into
\begin{equation}
	a_{t}+3(a+b)a_{x}+a_{xxx}+b_{t}+3(a+b)b_{x}+b_{xxx}=0, \label{abeq}
\end{equation}%
where the first three terms of this equation are anti-$\mathcal{PT}$-symmetrically related to the last three terms. Splitting then (\ref{abeq}) accordingly, one may also add an arbitrary $\mathcal{PT}$-invariant function, i.e. $
\mathcal{PT} f(a,b) = f(a,b)$, to the first and subtract it from the second equation. This function constitutes an ambiguity that vanishes when one adds the two equations back together again. As will be discussed below, different choices lead to different types of behaviour and also affect the stability of the resulting system.

Here we select two particular cases. Taking the function $f$ for instance to be $f_1(a,b)=3/2(a+b)(b_{x}-a_{x})$ one obtains the two  equations%
\begin{eqnarray}
	a_{t}+3/2(a+b)(a_{x}+b_{x})+a_{xxx} &=&0,  \label{AB1} \\
	b_{t}+3/2(a+b)(a_{x}+b_{x})+b_{xxx} &=&0,  \label{AB2}
\end{eqnarray}
whereas the choice $f_2(a,b)=3/4(a+b)(b_{x}-a_{x})$ produces
\begin{eqnarray}
	a_{t}+3/4(a+b)(3a_{x}+b_{x})+a_{xxx} &=&0,  \label{AB12} \\
	b_{t}+3/4(a+b)(a_{x}+3b_{x})+b_{xxx} &=&0.  \label{AB22}
\end{eqnarray}
The equations in each pair (\ref{AB1}), (\ref{AB2}) and  (\ref{AB12}), (\ref{AB22}) are anti-$\mathcal{PT}$-symmetrically related to each other. They are often referred to as Alice and Bob KdV (AB-KdV) equations in reference to conventions used in discussions within the context of cryptographic systems. 

Alternatively we may also obtain the AB-KdV equations (\ref{AB1}) and (\ref{AB2}) from a Hamiltonian with $\mathcal{PT}$-invariant Hamiltonian density 
\begin{equation}
	{\cal{H}} = a_x b_x -\frac{1}{4} (a+b)^3
	\label{Ham1AB}
\end{equation}
when using functional variations for a real  Hamiltonian with two independent fields $a$ and $b$ in the form
\begin{eqnarray}
	a_t &=& \partial_x \frac{\delta H }{\delta b} = \partial_x \left(  \frac{\partial {\cal{H}}}{ \partial b} - \frac{d}{dx} \frac{\partial {\cal{H}}}{ \partial b_x}
	+ \frac{d^2}{dx^2} \frac{\partial {\cal{H}}}{ \partial b_{xx}}+ \ldots \right) , \\
		b_t &=& \partial_x \frac{\delta H }{\delta a} = \partial_x \left(  \frac{\partial {\cal{H}}}{ \partial a} - \frac{d}{dx} \frac{\partial {\cal{H}}}{ \partial a_x}
	+ \frac{d^2}{dx^2} \frac{\partial {\cal{H}}}{ \partial a_{xx}}+ \ldots \right).
\end{eqnarray}
When compared with previously proposed Hamiltonian structure for the nonlocal Hirota equation we note the difference here is due to the well-known bi-Hamiltonian structure of the KdV-equation \cite{Gardner,Faddeev,Das} that is not inherited by their AB-versions.

\subsection{Orbitally stability of nonlocal Alice and Bob solutions}
Let us now comment on the stability of the solutions $%
a(x,t)$ and $b(x,t)$ to these equations. We will treat the two systems simultaneously and also focus in each case just on one equation, (\ref{AB1}) and (\ref{AB12}), as the corresponding partner equations are simply obtained by a $\mathcal{PT}$-transformation. Using the travelling wave coordinate $\zeta$ converts the two equations into
\begin{equation}
	a_{\zeta \zeta \zeta} +\frac{4 \tau_1}{\alpha ^2} (a+b) \left(\tau_2 a_\zeta +b_\zeta\right) -4 a_\zeta =0, \label{abtravel}
\end{equation}
with $\tau_1= 3/2$, $\tau_2=1$ for $f_1$ and $\tau_1= 3/4$, $\tau_2=3$ for $f_2$. Next we perturb both solutions for each of the pair of equations linearly as
\begin{equation}
	a(\zeta) \rightarrow  	a(\zeta) + \epsilon \sigma_a(\zeta), \qquad b(\zeta) \rightarrow  	b(\zeta) + \epsilon \sigma_b(\zeta), \qquad \epsilon \ll 1. \label{abp12}
\end{equation}
so that the first order equation in $\epsilon$ becomes
\begin{equation}
\frac{4 \tau _1 \left(\sigma _a+\sigma _b\right) \left(\tau _2 a'+b'\right)}{\alpha ^2}+ \left[\frac{4 \tau
	_1 \tau _2 (a+b)}{\alpha ^2}-4\right] \sigma _a' +\frac{4 \tau _1 (a+b) }{\alpha ^2} \sigma _b' +\sigma_a'''=0 . \label{firsto}
\end{equation}

Assuming that the two perturbations are anti-$\mathcal{PT}$ related to each other, i.e. $\sigma_b(\zeta)= -\sigma_a(\zeta)$, equation (\ref{firsto}) reduces to a Sturm-Liouville equation for $(\sigma_a)_{\zeta}$  
\begin{equation}
	\left[ (\sigma_a)_{\zeta}  \right]_{\zeta \zeta}+ \frac{8 \tau_1 (\tau_2-1)}{\alpha ^2} u (\sigma_a)_{\zeta}  = 4 (\sigma_a)_{\zeta}, \label{SLsig}
\end{equation}
for the potential $V = [8 \tau_1 (\tau_2-1)]/ \alpha^2 u$. Moreover, all higher order terms in $\epsilon$ vanish, so that for the solutions of (\ref{SLsig}) the combinations in (\ref{abp12}) always constitute new solutions to the AB-KdV equation with a free parameter $\epsilon \rightarrow \nu$, rather than a perturbation thereof. In order to distinguish this behaviour from linear perturbations with nonvanishing higher order terms, we refer to this behaviour here as orbital stability in a slight abuse of language when compared to the usage of the term in the context of functional analysis.

\subsubsection{Unstable solutions for the type I system}
We note that for the $f_1$ choice, $\tau_1= 3/2$, $\tau_2=1$, the potential term in the auxiliary Sturm-Liouville equation (\ref{SLsig}) drops out so that the equation is simply solved by
\begin{equation}
\sigma_a(\zeta)  = c_3+\frac{1}{2} \left(c_1 e^{2 \zeta }-c_2 e^{-2 \zeta }\right) .
\end{equation}
Thus, apart from the trivial solution $c_1=c_2=0$, $\sigma_a(\zeta)$ is asymptotically divergent for either $x \rightarrow \infty$ or  $x \rightarrow -\infty$ and also  $t \rightarrow \infty$ or  $t \rightarrow -\infty$. Thus {\em any} solution for this type of AB-KdV-equation is unstable. We find a similar behaviour when assuming the two perturbations to be $\mathcal{PT}$ related to each other, i.e. $\sigma_b(\zeta)= \sigma_a(\zeta)$, which we, however, do not report here.  

\subsubsection{Orbitally stable solutions for the type II system}

In contrast, we may find orbitally stable solutions for the second type of system with the choice of $\tau_1= 3/4$, $\tau_2=3$ for $f_2$. First we notice that for any given solution $u(x,t)$ of the KdV equation we may set $a(x,t)=b(x,t)=u(x,t)$ and construct a new solution by means of (\ref{SLsig}) and (\ref{abp12}) since all higher order terms in $\epsilon$ vanish. For instance, starting with the trivial solution $u(x,t)=c$, we easily solve (\ref{SLsig}) and obtain the solution
\begin{equation}
	a(\zeta)=c +  \nu \frac{\alpha}{\sqrt{\alpha ^2-3 c}}  \sinh \left(\frac{2   \sqrt{\alpha ^2-3 c}}{\alpha } \zeta \right), \quad 	b(\zeta)=c -  \nu \frac{\alpha}{\sqrt{\alpha ^2-3 c}}  \sinh \left(\frac{2   \sqrt{\alpha ^2-3 c}}{\alpha } \zeta \right), \label{statsol}
\end{equation}
which is oscillatory when $\alpha^2 < 3 c$ but asymptotically divergent for $\alpha^2 > 3 c$. In the first instance this will allow us to draw a conclusion about the stability of the solution $a(x,t)=b(x,t)=c$, however, noting that in (\ref{SLsig}) only $u$ occurs, these stability properties are also shared with the solutions in (\ref{statsol}). Notice that in both regimes the solutions are related via ${\cal PT}$-symmetry, so that one should not draw the conclusion that the stability is governed by the symmetry. 

\begin{figure}[h]
	\centering         
	\begin{minipage}[b]{0.5\textwidth}           \includegraphics[width=\textwidth]{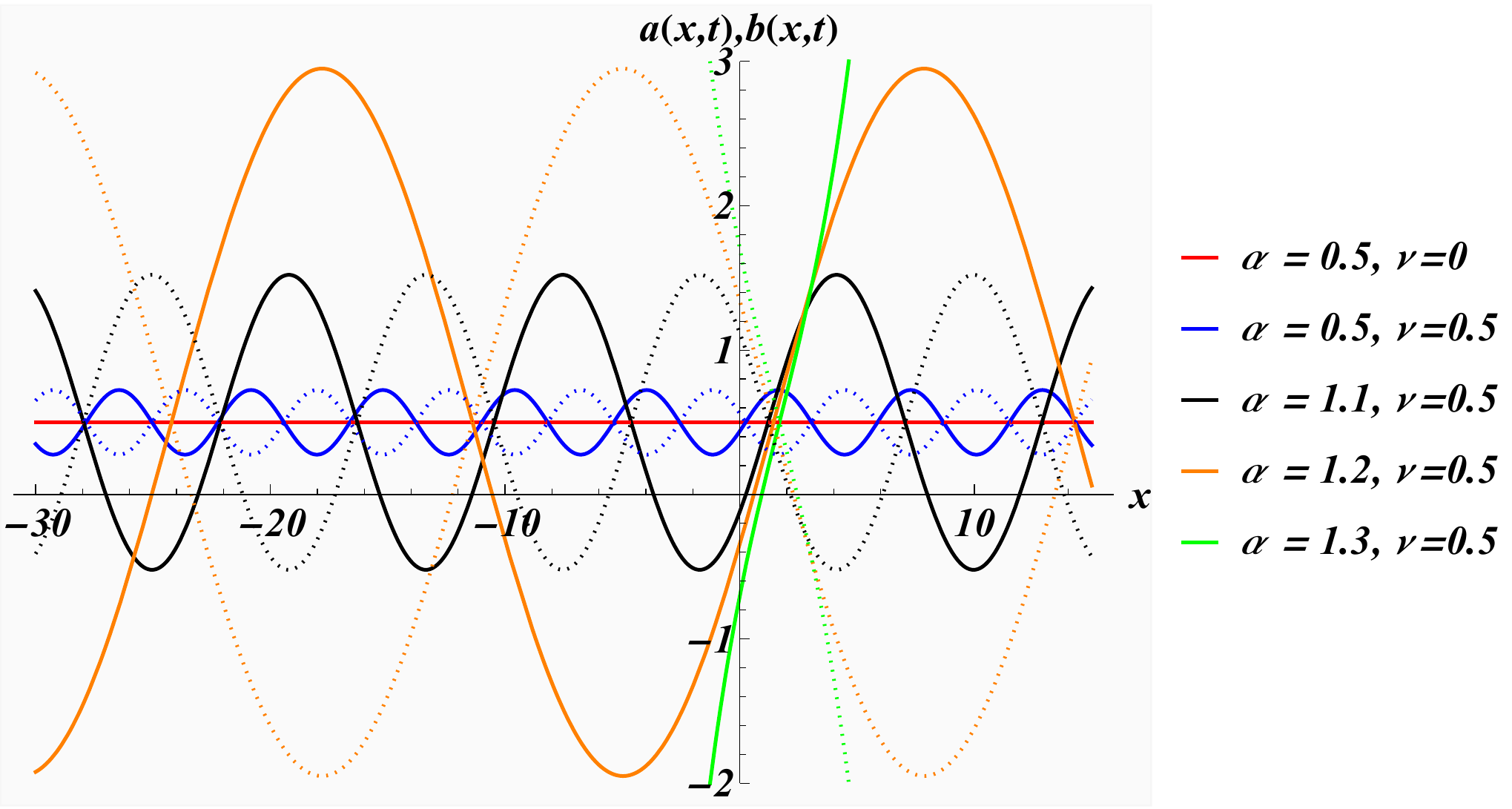}
	\end{minipage}   
	\begin{minipage}[b]{0.45\textwidth}           
		\includegraphics[width=\textwidth]{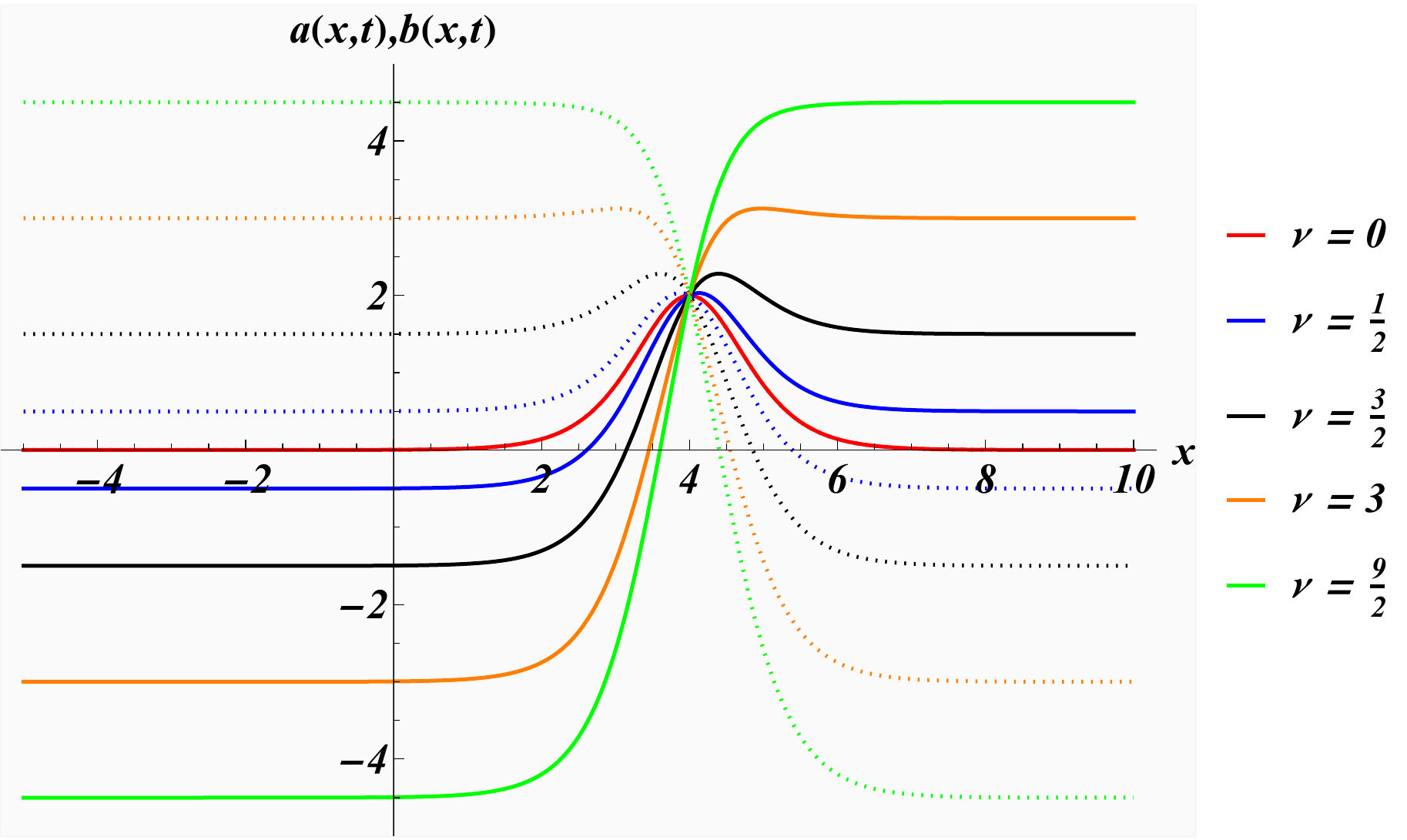}
	\end{minipage}    
	\caption{Panel (a): Orbitally stable and unstable solutions solutions $a(x,t)$ (solid) and $b(x,t)$ (dotted) in (\ref{statsol}) to the AB-KdV equation with $c=1/2$,  $\alpha_c=\sqrt{3/2} \approx 1.22$ at $t=1$ for different values of $\alpha$ and $\nu$. Panel (b): Orbitally stable solutions $a(x,t)$ (solid) and $b(x,t)$ (dotted) in (\ref{absolution}) to the AB-KdV equation with $\alpha=2$ at $t=1$ for different values of $\nu$.}
	\label{abkdvkinksol}
\end{figure}

Taking $u(x,t)$ instead in the form (\ref{ones}), as a particular solution for the stable AB-KdV equation we solve (\ref{SLsig}) to
\begin{equation}
	(\sigma_a)_{\zeta} = \sech^2(\zeta) \quad \Rightarrow \quad  \sigma_a(x,t) = \tanh\left[ \frac{1}{2} \left(\alpha  x-\alpha ^3 t\right) \right] , \label{sig}
\end{equation}
so that we obtain the AB-KdV solution 
\begin{equation}
	a(x,t) = u(x,t) + \nu \tanh\left[ \frac{1}{2} \left(\alpha  x-\alpha ^3 t\right) \right], \qquad b(x,t) = u(x,t) - \nu \tanh\left[ \frac{1}{2} \left(\alpha  x-\alpha ^3 t\right) \right],   \label{absolution}
\end{equation}
which is stable by the same reasoning as above. We depict the solutions (\ref{statsol}) and (\ref{absolution}) together with their orbital perturbations in figure \ref{abkdvkinksol}. 

In panel (a) we see how the constant solution converts into oscillatory solutions until it becomes unstable when the critical value $\alpha_c = \sqrt{3c}$ is passed. Panel (b) depicts how the initial soliton solutions is gradually deformed into kink and anti-kink solutions with asymptotic values $\pm \nu$.

\section{Alice and Bob Boussinesq system}
\begin{figure}[h]
	\centering         
	\begin{minipage}[b]{0.52\textwidth}           \includegraphics[width=\textwidth]{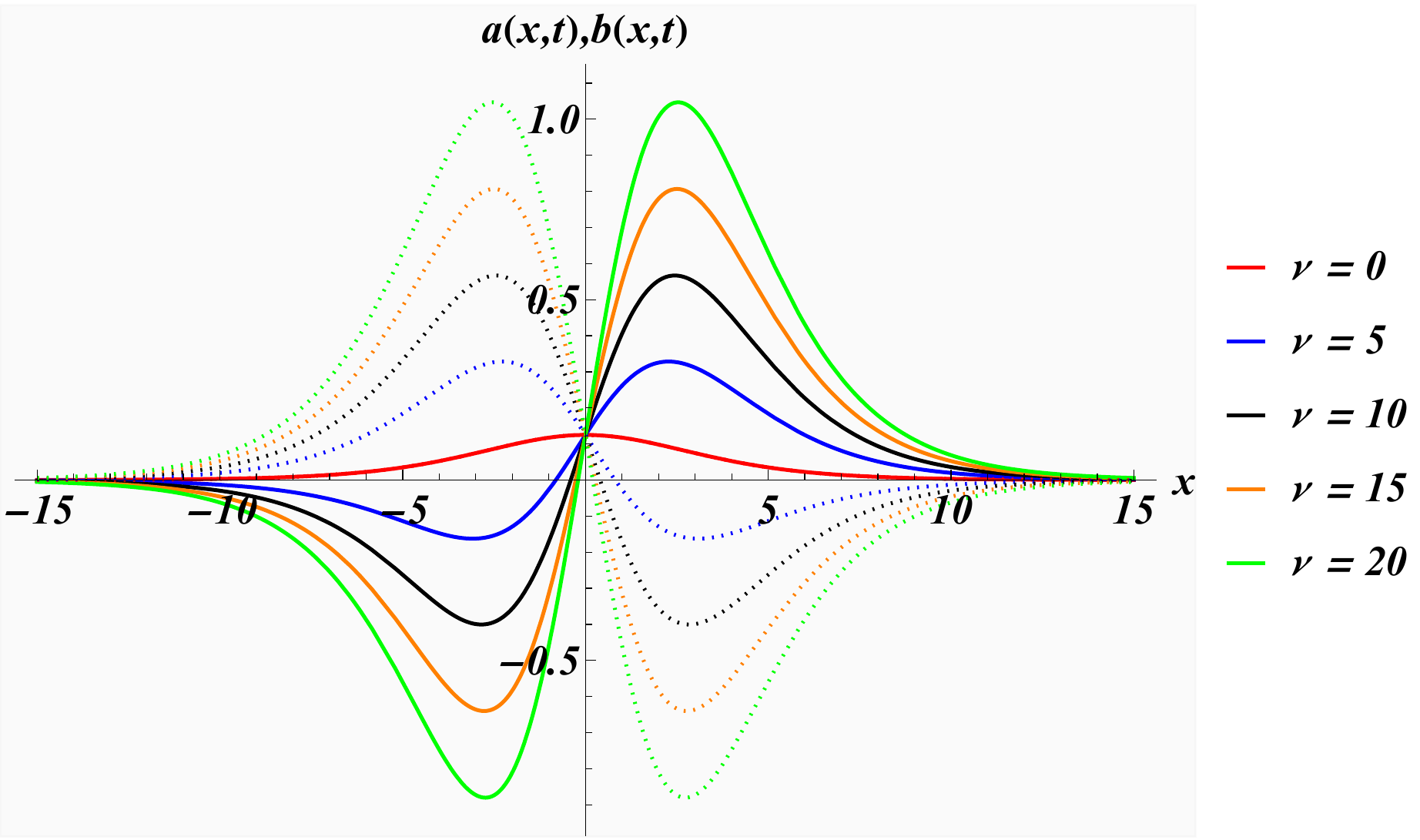}
	\end{minipage}   
	\begin{minipage}[b]{0.44\textwidth}           
		\includegraphics[width=\textwidth]{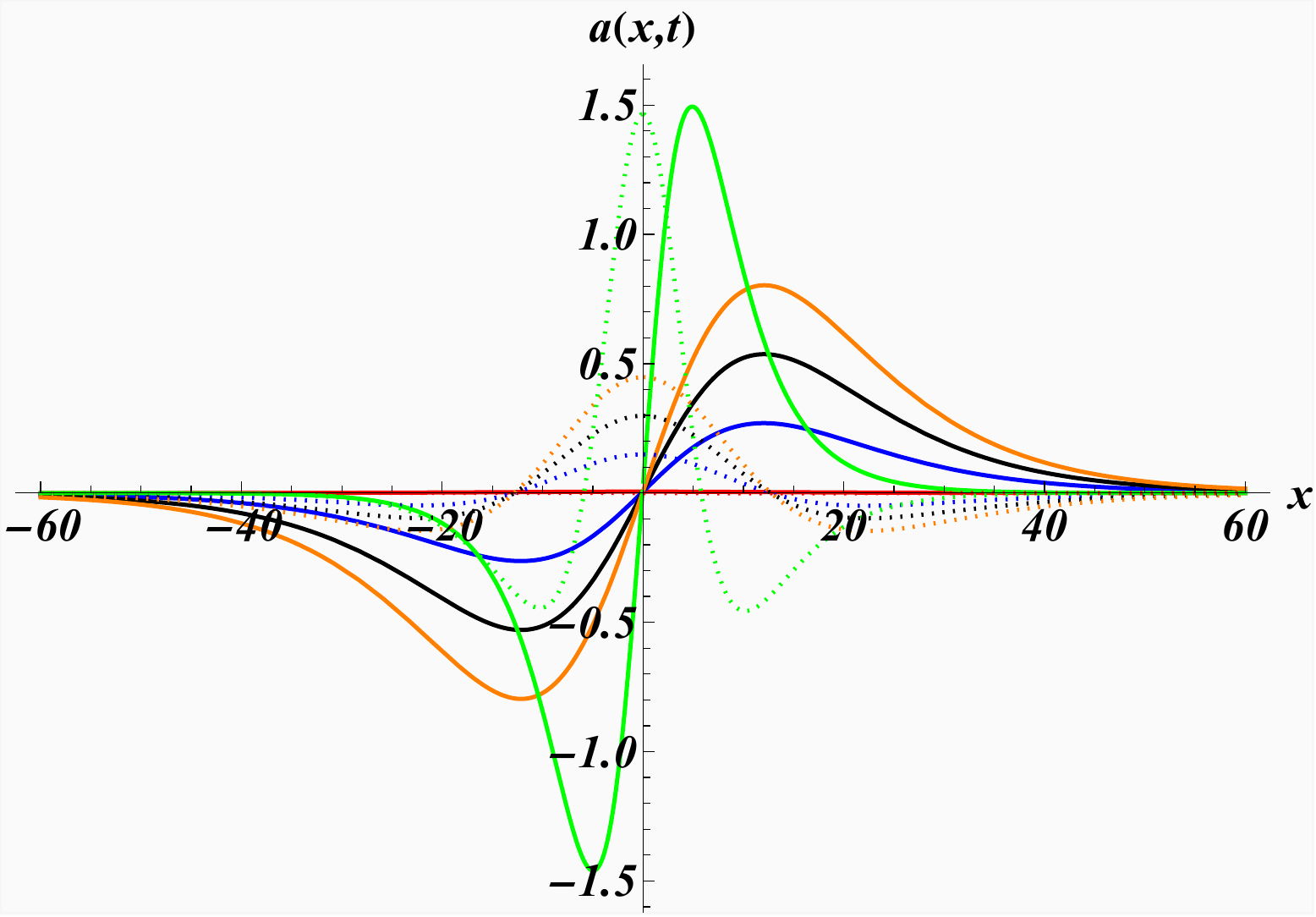}
	\end{minipage}    
	\caption{AB-Boussinesq solutions (\ref{bouab}) for different values of $\nu$. Panel (a) For $t=0$, $\alpha =1/2$, $\delta = 2$ with $a(x,t)$ (solid) and $b(x,t)$ (dotted). Panel (b) For $t=2$, $\alpha =1/10$, $\delta = -5$ with $ Re[a(x,t)]$ (solid) and $Im[b(x,t)]$ (dotted). }
	\label{Boupic}
\end{figure}

A further example of a so-called Alice and Bob system is a modification \cite{li2019multiple} of the Boussinesq equation 
\begin{equation}
	u_{tt}- \left( 3 u^2 + \delta u + u_{xx}   \right)_{xx} =0,  \qquad \delta \in \mathbb{R}. \label{Bou}
\end{equation}
Using the same approach as described for the KdV system by separating $u(x,t)$ into a sum of two $\mathcal{PT}$ related functions one derives \cite{li2019multiple} the following nonlocal AB-Boussinesq equation  
\begin{equation}
	a_{tt}- \left(  \frac{9}{4} a^2 + \frac{3}{2} a b -  \frac{3}{4} b^2  +\delta a +  a_{xx}   \right)_{xx} =0, \label{ABbous} 
\end{equation}  
where $b(x,t) = \mathcal{PT} a(x,t) $. Thus when perturbing the fields as
\begin{equation}
	a(x,t) \rightarrow  	u(x,t) + \epsilon \sigma(x,t), \qquad b(x,t) \rightarrow  	u(x,t) - \epsilon \sigma(x,t), \qquad \epsilon \ll 1, \label{abp}
\end{equation}
the zeroth order in $\epsilon$ becomes the Boussinesq equation and the first order equation reads
\begin{equation}
 \sigma _{{tt}}-12 u_x \sigma _x-6 \sigma  u_{{xx}}-6 u \sigma _{{xx}}-\delta  \sigma _{{xx}}-\sigma_{{xxxx}}=0. \label{Boufirst}
\end{equation}
Similarly as in the previous subsection, all higher order equations are zero. Equation (\ref{Boufirst}) is solved by taking $\sigma = c_1 u_x + c_2 u_ t$ for arbitrary constants $c_1$, $c_2$ as it converts (\ref{Boufirst}) into
\begin{equation}
 (c_1 \partial_x + c_2 \partial_t)	\left[u_{tt}- \left( 3 u^2 + \delta u + u_{xx}   \right)_{xx} \right]=0,  \qquad \delta \in \mathbb{R}. 
\end{equation} 
Thus starting from the concrete solution to the Boussinesq equation
\begin{equation}
u(x,t)= \frac{\alpha ^2}{1 + \cosh \left( \alpha  x \pm \alpha  \sqrt{\alpha ^2+\delta } t\right)}, \label{Bousol}
\end{equation} 
we obtain the new exact solution
\begin{equation}
	a(x,t) = b(-x,-t)= u(x,t) - \nu  \sinh ^4\left[\frac{1}{2} \left(\alpha  x\pm \alpha  t \sqrt{\alpha ^2+\delta }\right)\right] \text{csch}^3\left[\alpha 
	x\pm \alpha  t \sqrt{\alpha ^2+\delta }\right]. \label{bouab}
\end{equation}
We absorbed here the constants into the arbitrary parameter $\nu = 4 \epsilon \alpha^3 (c_1 \pm c_2 \sqrt{\alpha^2 +\delta})$. By the same reasoning as in the previous subsection we conclude that the solutions (\ref{Bousol}) and (\ref{bouab}) to the AB-Boussinesq equation are linearly stable. As also seen in figure (\ref{Boupic}) the solutions are regular, asymptotically vanishing real for $\alpha ^2+\delta \geq 0$ (panel a) and complex for $\alpha ^2+\delta < 0$ (panel b).  

\section{Conclusions}
We analysed several known one-soliton solutions with regard to their stability for different types of nonlocal nonlinear equations. For the nonlocal version of the Hirota system that contains interaction terms of fields at a particular space-time point with their complex parity conjugates, we found that the standard oscillatory and nonstandard rogue wave solutions have real energies and are stable. The AB-KdV equation can be formulated in different versions. For the most natural symmetric version, referred to here as type I, we identified a Hamiltonian, but showed that all solutions to these equations are unstable with regard to the perturbations analysed. In contrast, the type II AB-KdV equation possess solutions that are stable in a certain parameter regime, but become unstable when a critical value is passed. The AB-Boussinesq is also shown to possess stable solutions.

There are a number of interesting open questions left for further investigation. All investigated models allow for different versions of nonlocality. As discussed in \cite{CenFringHir}, the nonlocal Hirota equation allows for variants that besides $q(x,t)$ also contains the fields $q^\ast(-x,t)$, presented here, or  $q^\ast(x,-t)$,  $q^\ast(-x,-t)$, $q(-x,t)$, $q(x,-t)$,  $q(-x,-t)$. The last five possibilities have not been treated here, but given that these models exhibit quite distinct types of behaviour, it would be very interesting to settle as well the stability status of their solutions. Another open, and challenging issue is to clarify whether the multi-soliton solutions are stable or unstable.

\medskip

\noindent \textbf{Acknowledgments:} JC is supported by the U.S. Department of Energy through the LANL/LDRD Program and the Center for Nonlinear Studies. FC was partially supported by Fondecyt
grant 1211356. TT is supported by EPSRC grant EP/W522351/1. 

\newif\ifabfull\abfulltrue

\newif\ifabfull\abfulltrue


\end{document}